# Computational Cannula Microscopy of neurons using Neural Networks

Ruipeng Guo,[1] Zhimeng Pan,[2] Andrew Taibi,[3] Jason Shepherd[3] and Rajesh Menon[1,*]

[1]Department of Electrical and Computer Engineering, University of Utah
[2]School of Computing, University of Utah
[3]Department of Neurobiology and Anatomy, Biochemistry and Ophthalmology and Visual Sciences, University of Utah
* rmenon@eng.utah.edu

**Abstract:**
Computational Cannula Microscopy is a minimally invasive imaging technique that can enable high-resolution imaging deep inside tissue. Here, we apply artificial neural networks to enable fast, power-efficient image reconstructions that are more efficiently scalable to larger fields of view. Specifically, we demonstrate widefield fluorescence microscopy of cultured neurons and fluorescent beads with field of view of 200μm (diameter) and resolution of less than 10μm using a cannula of diameter of only 220μm. In addition, we show that this approach can also be extended to macro-photography.

## Introduction

Deep tissue imaging with high resolution and low damage is vital in biological and in medical fields, especially for the study of neurons *in vivo*. Multi-photon microscopy and micro-endoscopy are popular approaches [1-3]. However, it is still challenging to achieve high resolution, low damage and deep imaging simultaneously. Besides, most of these systems require high-power coherent light sources and complex optical systems driving up overall costs. In 2013, we demonstrated the approach of using a surgical cannula in conjunction with LED illumination to achieve fluorescence microscopy inside brain tissue with close to diffraction-limited resolution, a technique we refer to as computational-cannula microscopy (CCM) [4,5]. The cannula behaves as a lightpipe, guiding excitation light inside the tissue. Subsequent fluorescence is then collected by the same cannula (analogous to an epi-configuration) and guided to the outside world. Due to the incoherence of the fluorescence and due to the excitation of multiple guided modes within the cannula, the output signal has little resemblance to the input. Previously, we showed that via painstaking calibration of the system, we can apply regularization-based linear-algebraic techniques to obtain the reconstructed image. We emphasize that this approach is in contrast to alternative computational imaging approaches because we do not require temporal coherence in the excitation nor in the emission [4-6]. Our approach is a snap-shot microscopy technique, and no scanning is required. However, major disadvantages of CCM are the requirement for slow calibration and the relatively high computation time. Furthermore, photobleaching of the fluorescent beads used for calibration restrict the achievable resolution and field of view. In this Letter, we show that by training an artificial neural network (ANN), we can overcome these limitations. Specifically, we utilized this trained ANN to achieve image reconstructions of cultured neurons with < ~1% maximum average error relative to the reference image, resolution of ~5μm and field of view of ~200μm. Most importantly, the reconstruction time is only weakly dependent on the image size, allowing for scaling the field of view and resolution in the future. We present only *ex vivo* results here and *in vivo* imaging will be the subject of future work. In addition, we show that cannula-based imaging can be extended to macro-photography. This is demonstrated by imaging a liquid-crystal display using the cannula and computationally reconstructing the images via a trained ANN.

## Experiment

The schematic of our Computational Cannula Microscope (CCM) is shown in Fig. 1. The cannula (see photograph in the inset) is made by removing the sheath of the fiber (FT200EMT, Thorlabs) and placing it in a stainless steel ferrule(SFLC230, Thorlabs). The cannula diameter and length are 220μm and 8mm, respectively [4]. Excitation from a blue LED (center wavelength = 470nm, M470L3, Thorlabs) is conditioned and focused onto the top face of the cannula via a 20X objective (PLN 20X, Olympus). The cannula guides the excitation to its bottom face and uniformly illuminates a sample placed in close proximity (typical gap ~ 200μm). The field and collector lenses are adjusted to ensure that the excitation region is as uniform as possible. The fluorescence from the sample is collected by the same cannula and guided to its top face, which is then imaged onto a sCMOS camera (C11440, HAMAMATSU). We set up a 520nm-35nm filter and a 472nm-30nm filter in the optical path to separate the fluorescent signals from the source beam. An exemplary image is shown as inset. A reference widefield fluorescence microscope was built to image the same sample from underneath as illustrated in Fig. 1. The objectives of both microscopes have the same magnification. The corresponding image collected by the reference scope is also shown in the inset. The field of view (diameter of circle in the CCM image) is 200μm and that of the reference microscope is ~260μm.

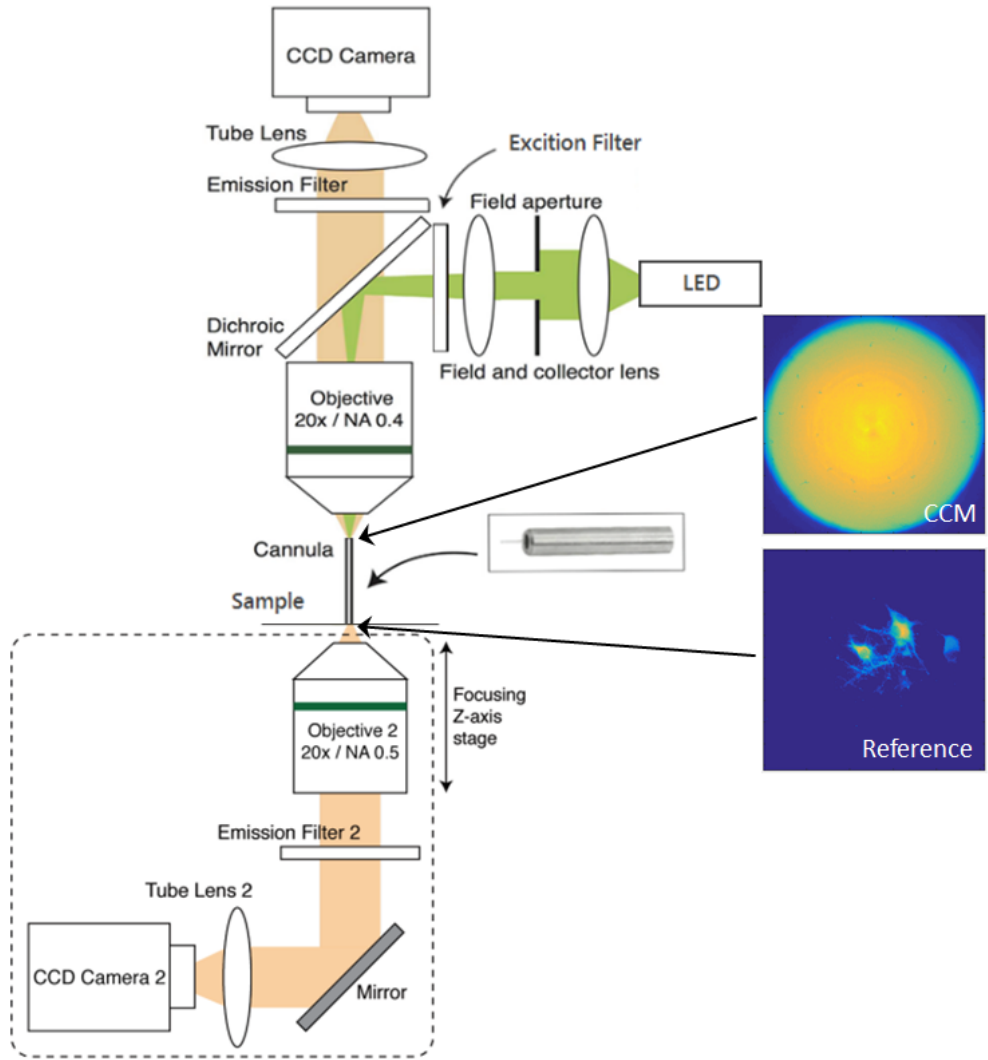

Fig 1: Schematic of our experimental system. The CCM is above the sample, while the reference widefield microscope is below the sample delineated by dashed lines. A photograph of the cannula is shown for reference. The other insets illustrate an example fluorescence signal on the top face of the cannula, and the corresponding reference image collected by the reference microscope.

*Primary Neuron Culture:* Primary neurons were taken from dissociated hippocampi of E18.5 Sprague-Dawley rat pups. Hippocampi were dissociated using 0.01% DNase (Sigma-Aldrich) and 0.067% papain (Worthington Biochemicals) prior to trituration through glass pipettes to obtain a single-cell suspension. Cells were then plated at 8 x$10^4$ cells/ml in Neurobasal medium (Thermo-Fisher) supplemented with 5% horse serum, 2% GlutaMax (Thermo Fisher), 2% B-27 (Thermo Fisher), and 1% penicillin/streptomycin (Thermo Fisher) on coverslips (No. 1, Bioscience Tools) coated overnight with 0.2mg/ml poly-L-lysine (Sigma-Aldrich) in 100mM Tris-base (pH 8). Neurons were grown at 37°C/5% CO and fed via half-media exchange every 3$^{rd}$ day with astrocyte conditioned Neurobasal media supplemented with 1% horse serum, 1% GlutaMax, 2% B-27, and 1% penicillin/streptomycin with the first feeding containing 5µM β-D-arabinofuranoside (Sigma-Aldrich) to limit overgrowth of glial cells. Neurons were grown for 12-14 days *in vitro* prior to transfection, fixation, and imaging.

*Neuron Transfection:* Neurons were transfected after 12 days *in vitro* with 0.5 µg of pCAG-eGFP (Addgene: 89684) using lipofectamine 2000 at a 3:1 ratio when complexed with plasmid DNA. Neurons were transfected over the course of 1 hour at 37°C in pH 7.4 Minimum Essential Media (Thermo Fisher) supplemented with 2 % GlutaMax, 2% B-27, 15mM HEPES (Thermo Fisher), 1mM Sodium Pyruvate (Thermo Fisher), and 33 mM Glucose. After transfection the neurons were given 24 hrs in growth media at 37°C/ 5% $CO_2$ to allow sufficient recovery and expression of the plasmid prior to fixation in 4% formaldehyde (thermo fisher)/4% sucrose (VWR) in phosphate buffered saline for 15 minutes at room temperature. After fixation neurons were mounted in Prolong Gold Aqueous Mounting Medium (Thermo Fisher) and imaged.

**Neural Network architecture**

The inverse problem that we are attempting to solve in CCM is an ill-conditioned linear system of equations that can be represented approximately as y = b*x + c, where y is the recorded sensor image, b is the system transfer function, x is the unknown object and c is the noise. ANNs have been shown to be good candidates to solve poorly conditioned inverse problems such as these previously [8]. Specifically, the universal approximation theorem guarantees that an ANN is able to closely approximate any continuous function similar to the one outlined above [9]. Secondly, the computational cost of an ANN is predictable and fixed, which is in contrast to regularization-based linear-algebraic techniques. Here, we apply a feed-forward ANN in the form of the well-known U-net modified with dense blocks from Res-net [10].

**Dataset**

The sample used to build the dataset is comprised of mGFP neurons on slides as described earlier. It is placed under the bottom face of the cannula as shown in Fig. 1. The distance between the slide and the cannula is about 200µm so that the sample slide can be moved in the horizontal plane without damage to the cannula. The sample is imaged by the reference microscope and the CCM, simultaneously. The images recorded by the reference microscope are used as the label images for training the ANN. To capture multiple images across the sample for creating the dataset, the slide is stepped using a stage in a raster fashion with the step size of 80µm. The field of view of this system is 200µm in diameter and the same neuron images were recorded at least four times. The vertical position of the stage is adjusted every 500 images, to keep all the neurons on the same plane, *i.e.*, in focus relative to the cannula.

     We acquired a total of 18,339 images. 16,504 images were randomly chosen for training and the remaining 1,835 images were used for testing. The size of reference images is 1,024*1,024 and that of the CCM image is 340*340. All images were first reshaped to either 128*128 or 256*256 to fit the ANN and also to speed up the training process for proof of principle. We trained a separate ANN for each case, which we refer to as ANN_128 and ANN_256, respectively.

     The ANN was trained by minimizing pixel-wise cross-entropy, which was previously shown to provide good results with reconstructions of sparse images [10]. The well-known ADAM optimizer was applied during training as has been described elsewhere [10].

## Results

Exemplary results for ANN_128 and ANN_256 are illustrated in Figs. 2a and 2b, respectively. In each panel, the CCM image, the reference image and the ANN output image are shown. We emphasize that these images are never seen by the trained ANN before testing. Clearly, both ANNs are able to reconstruct the images with good fidelity. We emphasize that the average computation times were 3.5ms and 15ms for ANN_128 and ANN_256, respectively. In comparison, the linear-algebraic technique required about 100ms for 128*128 images and 490ms for 256*256 images. The computer hardware used was Intel(R) Core(TM) i7-4790 CUP with clock frequency of 3.60GHz, memory of 16.0GB and the associated GPU is NVIDIA GeForce GTX 970. As expected, the ANN approach is 1 or 2 orders of magnitude faster. Even more importantly, the computation time for the ANN is only weakly dependent on the image size, which will enable easier scaling to higher resolutions and fields of view in the future.

It is noted that the resolution of the NN output images need further improvement. This can be achieved by increasing the resolution of the reference microscope in the future.

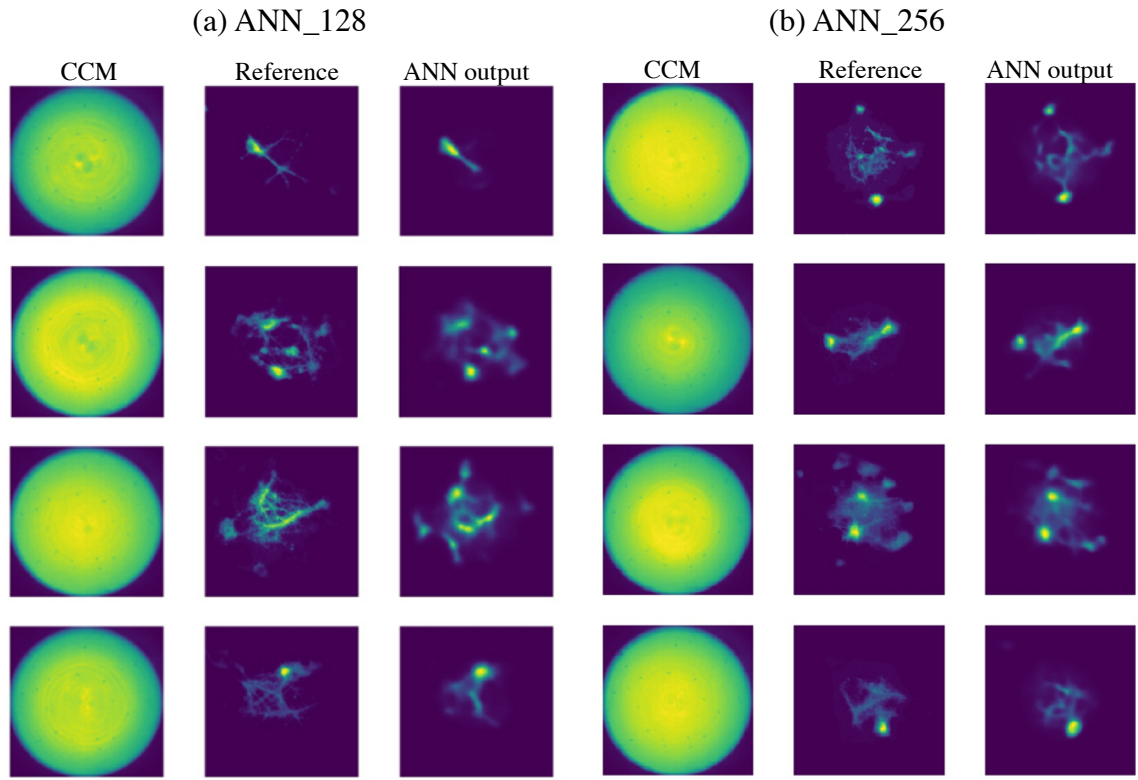

Figure 2. Imaging neurons with (a) ANN_128 and (b) ANN_256. In each panel, the raw sensor image (CCM), the reference image and the ANN output image are shown. The scale of each image is 200μm*200μm. We note that these images are not seen by the ANN during training.

The output images from each ANN were characterized using the structural-similarity index (SSIM) and the mean absolute error (MAE). The structural similarity (SSIM) index is a method to assess the similarity between the reconstructed image and the reference image, so the higher the value the better. MAE is the average error between reconstructed image and the reference image and hence, lower the value the better. Table 1 summarizes average SSIM and MAE values for the 2 ANNs. Both ANNs exhibit excellent image reproduction with MAE of ~1% or lower.

Table 1. Quantitative performance of the ANNs averaged over 1,835 testing images.

|       | ANN_128 Training | ANN_128 Testing | ANN_256 Training | ANN_256 Testing |
|-------|------------------|-----------------|------------------|-----------------|
| SSIM  | 0.9269           | 0.9290          | 0.9548           | 0.8933          |
| MAE   | 0.0095           | 0.0091          | 0.0066           | 0.0169          |

In order to clarify the resolution of our approach, we created a third dataset using 4μm-diameter fluorescence beads (FluoSpheres[TM] sulfate, Invitrogen). This dataset was then used to train a new pair of ANNs again with input images of 128*128 and 256*256 as previously (11,092 images were used for training and 1,233 images for testing). The trained networks were then used to reconstruct images that were not included in the testing phase. One example image with closely spaced beads was then used to estimate the resolution as summarized in Fig. 3. The reference image is included for comparison. The distance between closely spaced beads is ~7μm for ANN_128 and ~11μm for ANN_256, respectively. The full-width at half-maximum of the cross-section through a single bead is ~5μm for both networks.

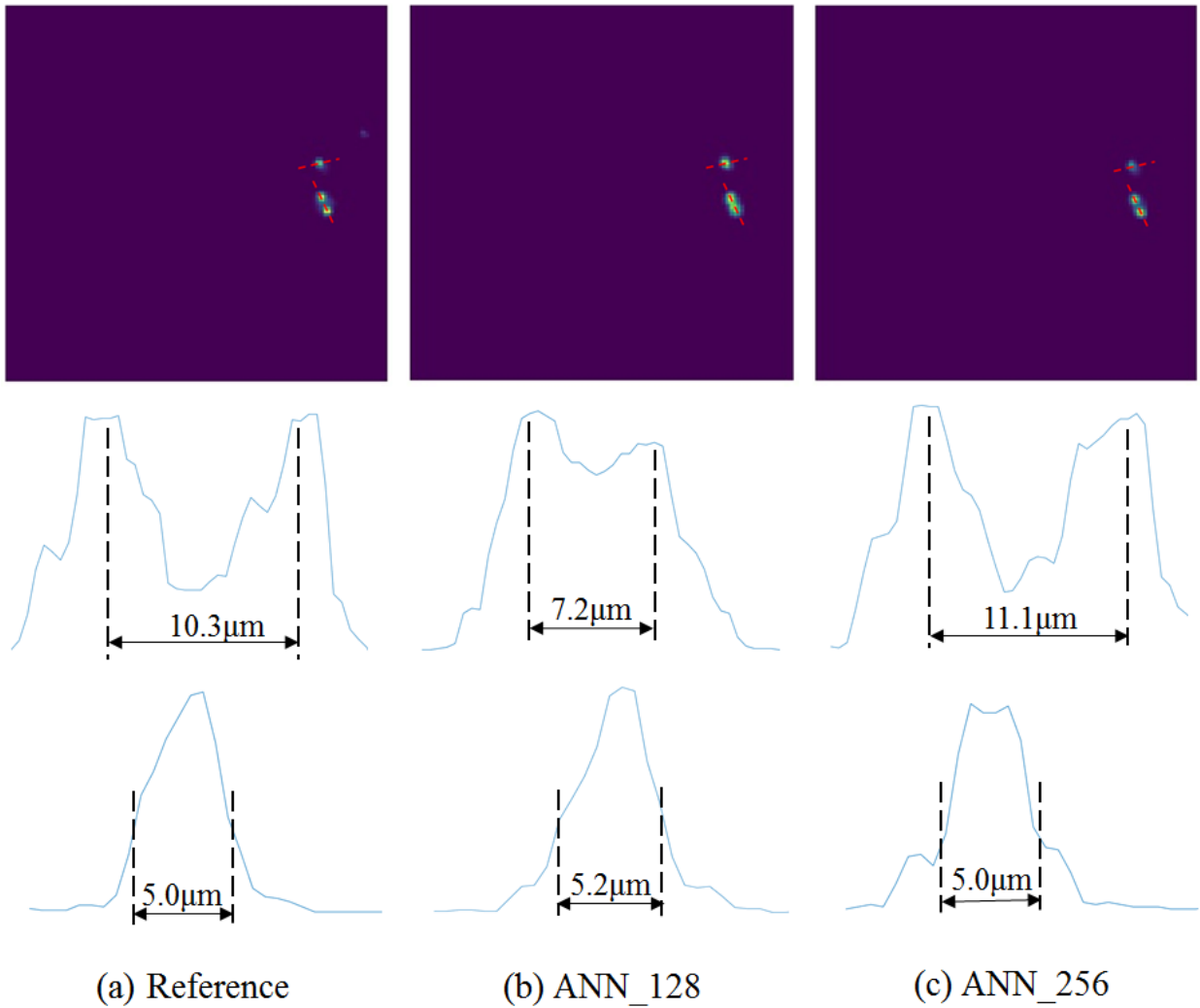

(a) Reference       (b) ANN_128       (c) ANN_256

Figure 3. Estimating resolution. Images of 4μm-diameter fluorescent beads: (a) Reference image, (b) Output of ANN_128 and (c) Output of ANN_256. Middle: Cross-section through neighboring beads. Bottom: Cross-section through one isolated bead. The full-width at half-maximum of the

ANN output suggests resolution of ~5μm. We note that these images are not seen by the ANN during training.

**Cannula Photography**
It is possible to generalize our approach to macro-photography as well. Previous work had indicated the potential for fully optics-less imaging where the image reconstructions were performed algebraically [11,12] and also with machine learning [10,13]. Here, we show that it is possible to image through the cannula itself with a field of view that is far larger than that determined by the cannula diameter. The experimental setup is illustrated in Fig. 4a. A liquid-crystal-display (LCD) was used as the object on which various test images were displayed. A cannula (length=12.5mm, diameter=220μm, CFMC52L02, Thorlabs) was placed at a distance 35cm away from the LCD. A lens was used to relay the image formed on the back face of the cannula onto a CMOS image sensor (MU300, AmScope). The dataset was comprised of 20,000 Kanji49 images [14], 20,000 EMNIST images [15] and 20,000 quasi-QR codes (generated randomly). All images were down-sampled to 128*128 pixels. 56,000 images were randomly chosen for training and the remaining 4,000 used for testing. The ANN had the same architecture as ANN_128 described earlier.

Figure 4b summarizes exemplary results from the trained ANN (rightmost column) compared to the reference image (that is displayed on the LCD, middle column) and the raw cannula image (left-most column). The ANN is able to form good reconstructions of even relatively complex images. The size of the image on the LCD is 5.5cm*5.5cm. Therefore, the angular field of view of this camera is ~9 degrees. The average SSIM and MAE of this ANN over the 4000 testing images were 0.8075 and 0.0682, respectively.

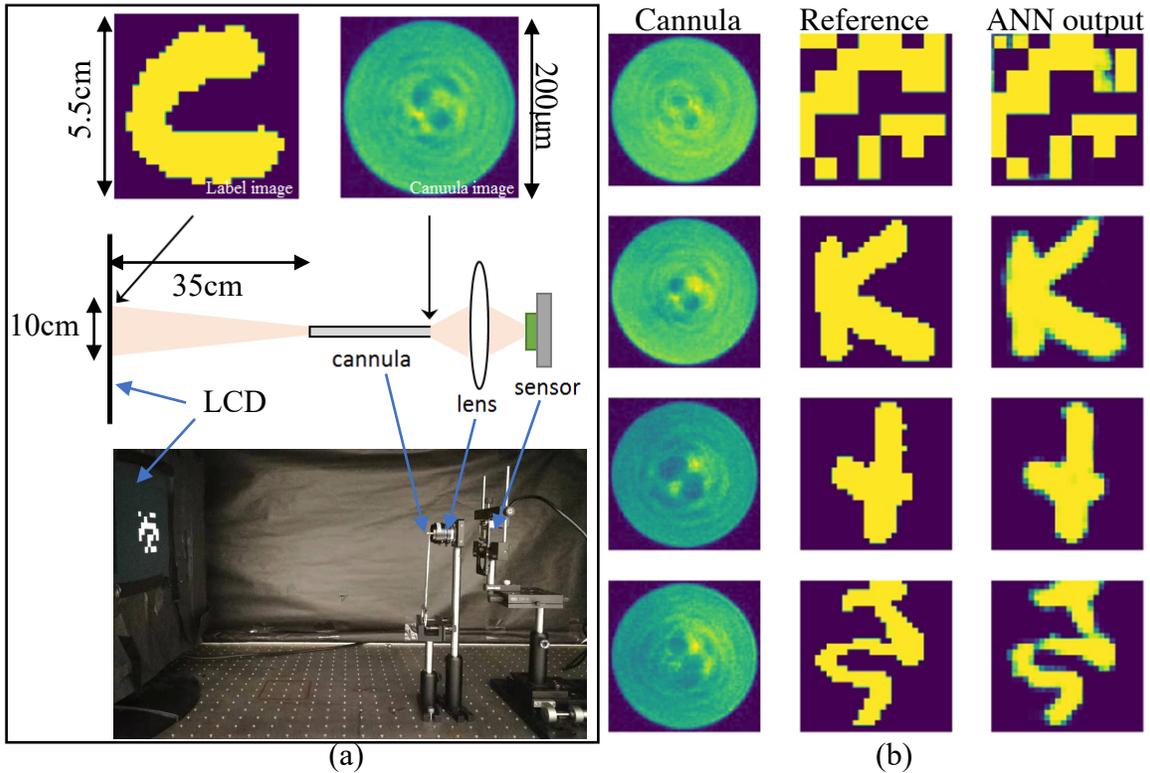

Figure 4. Cannula Photography. (a) Schematic of system. A cannula is placed facing an LCD. Test images were displayed on the cannula and the a lens relays the light intensity pattern on the right-face of the cannula to a CMOS image sensor. (b) Results after training an ANN with 128*128 pixel images. We note that these images are not seen by the ANN during training.

## Conclusions

Imaging is a means of information transfer from an object to an image sensor. Conventional imaging assumes optical systems that performs a one-to-one mapping of an object point to an image point. Recent work in the field of computational imaging is starting to expand this idea to include one-to-many mappings, where the information from one object point is mapped onto multiple image points. The information may then be recouped for human consumption via computational methods. In this Letter, we exploit these advances towards a vision of minimally invasive deep tissue imaging using a simple surgical cannula (whose diameter can be less than 200μm). The cannula performs two functions: guiding the excitation light to the region of interest and collecting the emitted fluorescence and guiding it to the outside world. Here, we trained two ANNs to reconstruct the images of the regions of interest with ~2 orders of magnitude improvement in the computation speed compared to linear-algebraic methods. In *ex vivo* imaging, we demonstrated resolution of about 5μm and field of view of 200μm. The resolution may be improved by: (1) increasing the network complexity (for instance, going to deeper networks, which will increase training complexity or via generative adversarial networks [16]), (2) by exploring optimal cannula geometries (this was briefly studied in ref [7]), and by increasing the resolution of the reference microscope. Although previously we have shown that CCM that is calibrated in air can be used *in vivo* [4], this is still an open question with ANNs. Finally, we point out that it is important to not bend or distort the cannula as this will create mode-mixing and could potentially invalidate the training of the ANN. However, the robustness of the ANN to such distortions is also an open question. Our final experiment that demonstrated a preliminary version of cannula macro-photography has important implications for imaging in hard-to-reach places like oil pipelines. Further work is required to understand its potential for color and depth imaging. Nevertheless, our experiments clearly indicate that computational methods, particularly neural networks can, not just complement optics, but could potentially even replace them for many applications.

**Funding:** National Science Foundation **(**10037833**)**

**Acknowledgements:** We thank Ganghun Kim for assistance with an earlier version of CCM.